# Mixed‒UNet: Refined Class Activation Mapping for Weakly-Supervised Semantic Segmentation with Multi-scale Inference

Yang Liu, Ersi Zhang, Lulu Xu, Chufan Xiao, Xiaoyun Zhong, Lijin Lian, Fang Li, Bin Jiang, Yuhan Dong, Lan Ma, Qiming Huang, Ming Xu, Yongbing Zhang, Dongmei Yu, Chenggang Yan, and Peiwu Qin

***Abstract*—** **Deep learning techniques have shown great potential in medical image processing, particularly through accurate and reliable image segmentation on magnetic resonance imaging (MRI) scans or computed tomography (CT) scans, which allow the localization and diagnosis of lesions. However, training these segmentation models requires a large number of manually annotated pixel-level labels, which are time-consuming and labor-intensive, in contrast to image-level labels that are easier to obtain. It is imperative to resolve this problem through weakly-supervised semantic segmentation models using image-level labels as supervision since it can significantly reduce human annotation efforts. Most of the advanced solutions exploit class activation mapping (CAM). However, the original CAMs rarely capture the precise boundaries of lesions. In this study, we propose the strategy of multi-scale inference to refine CAMs by reducing the detail loss in single-scale reasoning. For segmentation, we develop a novel model named Mixed-UNet, which has two parallel branches in the decoding phase. The results can be obtained after fusing the extracted features from two branches. We evaluate the designed Mixed-UNet against several prevalent deep learning-based segmentation approaches on our dataset collected from the local hospital and public datasets. The validation results demonstrate that our model surpasses available methods under the same supervision level in the segmentation of various lesions from brain imaging.**

## I. INTRODUCTION

IN recent years, deep learning techniques based on Convolutional Neural Networks (CNNs) have shown great potential in medical image processing [1-7]. Many advanced networks are adopted in the medical imaging field, such as locating lesions in magnetic resonance imaging (MRI) scans or computed tomography (CT) scans through semantic segmentation models, to assist doctors in diagnosing and treating diseases [8-10]. As an application in acute cerebral infarction (ACI), brain tissue necrosis caused by the sudden interruption of cerebral blood supply, the segmentation model can quickly detect the tiny and scattered cerebral infarcts in MRI scans (some infarcts are easily unnoticed by the naked eye) and pinpoint cerebral infarct boundaries, thereby avoiding a substantial expenditure of time and effort of experienced doctors and helping therapeutic decision-making [11, 12]. Unfortunately, deep learning applications based on lesion detection are limited because CNNs require many manually annotated images with pixel-level labels, which are time-consuming and labor-intensive. Plenty of works focusing on weakly-supervised semantic segmentation (WSSS) have been implemented to address this limitation and have achieved miraculous performance, even close to supervised learning [6, 7, 13-20]. These techniques rely on weaker forms of supervision, such as bounding boxes [16, 21, 22], points or squiggles [15, 23], image-level labels [18-20, 24-32], etc. Among them, image-level labels are widely adopted and followed in this work because they are easily obtained.

To localize lesions using only image-level labels, Zhou et al. propose class activation maps (CAMs) to excavate the most discriminative regions [33] by locating target objects where pixels with higher evaluation values are more likely to be within the target object. Since image-level labels only indicate whether the target objects exist, they do not contain any object location information [34]. Therefore, the localization ability of CAMs

This work was supported in part by Science, Technology, Innovation Commission of Shenzhen Municipality (JSGG20191129110812708; JSGG20200225150707332; ZDSYS202008201654000; JCYJ20190809180003689), National Natural Science Foundation of China (31970752), Shenzhen Bay Laboratory Open Funding (SZBL2020090501004), and China Postdoctoral Science Foundation (2020M680023).

Yang Liu, Bin Jiang and Dongmei Yu are with the Shandong University, Weihai 264209, China (e-mail: liuyangproven@mail.sdu.edu.cn; jiangbsdu@gmail.com; yudongmei198011@sina.cn ).

Fang Li, Lulu Xu, Chufan Xiao, Xiaoyun Zhong, Lijin Lian, Yuhan Dong and Lan Ma are with the Institute of Tsinghua-Berkeley Shenzhen, Tsinghua Shenzhen International Graduate School, Shenzhen 518000, China (e-mail: li.fang@sz.tsinghua.edu.cn; xululuuu@outlook.com; xiaocf21@mails.tsinghua.edu.cn; zhongxy21@mails.tsinghua.edu.cn; lianlijin2017@email.szu.edu.cn; Xi0218@hotmail.com; wmda130411@gmail.com ).

Ersi Zhang and Chenggang Yan are with the Hangzhou Dianzi University, Hangzhou 310000, China (e-mail: 15000852685@163.com; yanyg978@foxmail.com).

Qiming Huang and Ming Xu are with the Shenzhen ZNV Technology Co., Ltd, Shenzhen 51800, China (e-mail: qmhuang200123@163.com; mingy97876@outlook.com).

Yongbing Zhang is with the Shenzhen Graduate School of Harbin Institute of Technology, Shenzhen 51800, China (e-mail: yuanx20@mails.tsinghua.edu.cn).

Peiwu Qin is with Institute of Biopharmaceutical and Health Engineering, Tsinghua Shenzhen International Graduate School, Shenzhen 518000, China (e-mail: pwqin@sz.tsinghua.edu.cn).

can compensate for such image-level labelling problems, which further facilitates ill-posed weakly-supervised semantic segmentation under image-level supervision [32, 35]. However, the original CAMs are not enough to directly function as pixel-level supervision. The spatial resolution of the final convolutional layer output is low, causing the fused CAMs only to locate the coarse boundary of the lesion, limiting the accuracy of WSSS for lesion detection

Inspired by these issues, we propose a novel method to refine the CAMs by fusing the CAMs with multi-scale inference, capturing distinct and complementary features of objects. The model can capture the relative complete object features for small-scale objects, but the model can only perceive parts of the object for large-scale objects. Therefore, it can be inferred that the discriminative features captured by the model should be different for objects at different scales. For the same model and image, if the input image scales are different, the activation regions are also different. Thus, we propose that fusing CAMs with multi-scale achieves better localization performance [34, 36]. Nevertheless, Conditional Random Field (CRF) [37] is widely utilized as an independent post-processing process to refine original CAMs to match the ground truth as much as possible, which assigns the same labels to similar features by using color and position information. The refined CAMs we get after the application of CRF.

Pseudo masks can be obtained from the above refined CAMs, and these masks are utilized as pixel-level supervision to train a new segmentation method. Motivated by U-Net, commonly used in medical image segmentation, we design a Mixed-UNet by combining two U-Nets in a hybrid manner. In order to save more spatial details and reduce the computational complexity, Mixed-UNet has only three downsampling in the encoding stage, which is one layer less than U-Net. The encoding stage shares the same feature extractor, while the decoding stage is divided into two parallel branches with the same structure, effectively helping alleviate the lack of accuracy caused by a single branch in weakly-supervised segmentation on the condition of fewer parameters. Finally, the features extracted by the two branches are fused, and the final segmentation result is obtained through a softmax layer.

Here, we integrate the optimized CAMs as supervision into the Mixed-UNet to locate the lesions on our ACI dataset. The refined CAMs derive the pixel-level pseudo masks from image-level labels and are adopted to Mixed-UNet to segment lesions. Comparison of segmentation accuracy, computational time, and hardware resources with other prevalent intelligent semantic segmentation networks, our Mixed-UNet outperforms well. In addition, we use publicly available datasets associated with brain disease as validation sets to demonstrate that our model is a more effective and universal approach in the segmentation of lesions in brain imaging. The main contributions of this work are summarized as follows:

• We propose a method for avoiding enormous manual annotation in processing medical images by utilizing CAMs to excavate discriminative regions in a weakly supervised way, helping to locate the lesion only with image-level labels.

• Multi-scale CAMs are easy to build based on standard CNNs, making them have strong extensibility. Instead of single-scale inference, multi-scale CAMs can capture more complementary object features, thus reducing the loss of details and helping locate lesion boundaries more clearly.

• Our Mixed-UNet is a semantic segmentation network with two parallel branches in the upsampling phases, characterizing the architecture with robust scalability and flexibility. A series of empirical studies show that the Mixed-UNet achieves higher performance and is more computationally efficient than the popular brain lesion localization models.

## II. RELATED WORK

### A. Weakly-Supervised Semantic Segmentation

Semantic segmentation refers to classifying each pixel of an image as an instance, where each instance corresponds to a class. In the field of medical image processing, image segmentation can be used for image-guided intervention, radiotherapy, radiodiagnosis, etc. Different levels of supervision are available when training deep segmentation models, ranging from pixel-level annotations (supervised learning) and image-level and bounding box annotations (semi-supervised learning) to completely unannotated objects (unsupervised learning), where the last two levels of annotations belong to weak supervision [9, 38]. Training the architecture relies on a large amount of pixel-level labeled data, which is time-consuming and expensive, especially the pixel-level labels in medical images. However, a large number of images with the image-level label can be obtained in a relatively fast and inexpensive manner. Many weakly supervised semantic segmentation methods have emerged in recent years to alleviate the considerable burden of pixel-level annotation and have achieved miraculous performance, even close to supervised learning [6, 7, 13-20].

The general process of WSSS is conducted as follows: pixel-level pseudo-masks need to be generated by a weakly supervised algorithm at first. The images are then trained through a deep convolutional neural network. Finally, the output results and pseudo-masks are backpropagated to minimize the loss function and improve the model's performance. These techniques rely on weaker forms of supervision, such as bounding boxes [16, 21, 22], points or squiggles [15, 23], image-level labels [18-20, 24-32], etc.

Among them, the image-level label is the simplest form of weak labeling, which is relatively easy to obtain. Training images are labeled only by the classes they belong to, not by their location in the image. However, this also makes it challenging to use image-level labels to train segmentation networks, so many researchers start to consider building the correlations between image- and pixel-level labels.

Vezhnevets et al. propose a novel multi-image model (MIM) to recover the pixel labels of the training images based on the similarity of appearance [39]. In [40], Wei et al. make significant contributions to develop a flexible deep CNN infrastructure, where a shared CNN is connected. The output results are aggregated with max-pooling to produce the ultimate multi-label predictions. Based on such a flexible model, not only does the training network no longer need ground-truth bounding box information, but the network is robust to random noisy and redundant assumptions. Similar literature [41] explores a new framework consisting of two components. Reliable localization maps are first generated by combining hypothesis-aware classification and cross-image contextual refinement. Then, segmentation networks can be trained in a supervised manner from these generated localization maps. Besides, the authors explore two network training strategies to achieve good segmentation performance. The first strategy proposes a novel multi-label cross-entropy loss by directly using multiple localization maps to train the network, where each pixel contributes to each class with different weights. The second strategy infers a coarse segmentation mask from the localization map and uses the resulting mask to optimize the network based on a single-label cross-entropy loss. Kolesnikov et al. conduct CAMs to excavate the segmented objects, expand the regions according to the seeds, and get the segmentation results using CRF and boundary constraints [42]. Consequently, image-level labels are widely adopted and followed in this work due to their effortlessness to obtain.

### B. Class Activation Mapping

Weakly supervised semantic segmentation uses image-level labels to locate objects in images. Among them, some previous studies have proposed techniques of applying class activation mapping [33, 34, 43-45]. Whether it's CAM, Grad-CAM, or Score-Cam, they all follow a similar pipeline to generate CAMs. CNNs are trained corresponding to the classification objective at first, and then the CAMs are generated by global average pooling (GAP) on feature maps. Finally, the obtained seed region is thresholded based on the maximum value of the CAMs, and applied as weak supervision to the segmentation network to obtain the segmented target. These methods differ in the way of each feature map's weights generation. CAM [33] obtains weights from fully connected layers. Grad-CAM [43, 44] streams class-specific gradients to each feature map, and the gradient of each feature map is averaged as its weight. Whereas Score-CAM [45] eliminates the dependence on gradients and generates weights for each feature map through its forwarding score. They all generate reliable class activation maps from the final convolutional layer. Other works like CAM-GMP [46] and ACoL [28] add convolutional classification layers (Conv-Cls) on top of the backbone to generate CAMs directly to improve the integration and computational efficiency. In [47], the optimized dropout layer integrating erasure and discovery operations by attention mechanism improves the processing efficiency.

These approaches can identify the object regions by extracting the discriminative areas in class activation maps. However, due to the low spatial resolution of the output from the final convolutional layer, the generated CAMs can locate rough object regions, which cannot meet the accuracy requirements of weakly supervised semantic segmentation tasks. Significantly, the application performance in medical image processing is limited because the inherent characteristics of lesions are tiny and scattered [48].

To solve the problem of semantic segmentation of lesions, we turn our attention to some more advanced computer vision tasks, which benefit from the semantic knowledge of different feature scales [6, 49]. Shen et al. make full use of multi-scale features and fuse them for better tracking results [50]. LayerCAM is exploiting hierarchical semantic knowledge from different layers [34]. CAMs generated from shallow layers tend to capture fine-grained details of target objects, while CAMs generated from deep layers usually locate coarse spatial object regions, which means CAMs with multi-scale inherence all help to locate the lesions in target images.

### C. Encoder‑decoder semantic image segmentation networks

A fully convolutional network (FCN) is a CNN-based segmentation network firstly proposed by Long et al. [18]. It computes pixel-level outputs by upsampling the output activation maps. The context and spatial information in deep network images can be preserved by fusing the outputs with shallower layers' outputs. Encoder-decoder segmentation networks such as SegNet [4] are introduced based on FCN. The decoder network maps low-resolution encoder features to full-input resolution feature maps for pixel-level classification. The SegNet's decoder upsamples the lower-resolution input feature map. Specifically, the decoder performs non-linear upsampling using the pooling indices computed in the max-pooling step of the corresponding encoder. The architecture consists of a series of non-linear processing layers (encoders) and a corresponding set of decoder layers, followed by pixel classifiers. Typically, each encoder consists of one or more convolutional layers with batch normalization and ReLU non-linearity, followed by non-overlapping max-pooling and subsampling [51]. U-Net [2] consists of a contraction path to capture context and a symmetric expansion path to achieve precise localization. In U-Net, skip connections are added to the encoder-decoder image segmentation network to improve the model's accuracy and solve the problem of vanishing gradients. DenseNet [52] is a densely connected segmentation network architecture by adapting a U-Net-like encoder-decoder skeleton for more accurate segmentation. The feature fusing operation is modified using a spatial pyramid pooling module based on FCN. The spatial pyramid networks are able to encode multi-scale contextual information by probing the incoming features with filters or pooling operations at multiple rates and multiple effective fields-of-view, while the latter decoding networks can capture sharper object boundaries by gradually recovering the spatial information [6]. DeepLabV3+ combines the advantages of dilated convolution and feature pyramid pooling, adding a simple yet effective decoder module on top of DeepLabV3 to refine segmentation results, especially along object boundaries using dilated convolution and pyramid features, making it outperform many advanced segementation networks [7, 53, 54].

It is essential to compress the model depth to improve processing efficiency for medical image segmentation with larger volumes and resolutions, such as CT, MRI, and histopathology images. A neural architecture search method is applied to U-Net, and a smaller network with better organ/tumor segmentation performance is obtained on CT, MRI, and ultrasound images [55]. Brügger et al. redesign the U-Net architecture to make the network more memory efficient for 3D medical image segmentation by exploiting group normalization and the leaky ReLU function [56, 57]. Dilated Residual Network (DRN) has fewer parameters for medical image segmentation, making the architecture less prone to overfitting. Smaller networks with fewer parameters and higher efficiency, such a network structure is the future development direction of medical image segmentation [58].

## III. METHODS

We evaluate the proposed Mixed-UNet segmentation method and provide quantitative and qualitative analysis. The ACI dataset collected from the hospital is used for network training and the public database for validation. We first compare the efficiency of different CNN classification networks and choose the most efficient ResNet50 as the classification network. Based on ResNet50, multi-scale CAMs are applied to obtain pseudo-masks, which are applied as supervision in Mixed-UNet for image segmentation.

### A. Dataset

As shown in Figure 1, our ACI training dataset comprises 637 MRI scans from patients with ACI and 609 normal brain MRI slices from people without ACI, selected by experienced doctors. Among the slices with ACI, 310 slices are used as the training dataset and 133 slices as the testing dataset. Similarly, 427 slices are used for training, and 182 slices are used for testing among the normal slices. The splitting of the training and testing dataset coincides at the patient level, i.e., no brain slices from the same patient exist in training and testing datasets. The dataset is collected from the South University of Science and Technology Hospital in Guangdong Province with ethical approval and patients' notification.

The publicly available dataset about hemorrhage represents an independent dataset for validating the robustness and generalizability of our Mixed-UNet. Intracerebral hemorrhage (ICH) refers to sudden hemorrhage in brain tissue, ventricles, or both caused by rupture of a nontraumatic parenchymal blood vessel. This brain CT-hemorrhage dataset contains 100 normal and 100 hemorrhagic CT images from Kaggle. There is no difference in the type of bleeding. We resize the original CT slice into 256*256 and select 80 slices from the normal set and 80 slices from the hemorrhage as the training dataset. Then, 20 normal and 20 intracerebral hemorrhage slices are used as the test dataset.

### B. Medical Image Classification with CNNs

With the rapid development of deep learning, CNNs have led to a series of breakthroughs for image classification, which are able to extract significant features from medical images especially. The realization of CAMs is based on that the regions identified by CNNs can be highlighted for classification, that is, CAMs can be obtained by transforming high-efficiency and high-accuracy CNNs to extract lesion features from medical images. The traditional CNNs are mature, mainly consisting of convolution layers with learnable weights and biases, activation functions, pooling layers, and fully connected layers, conducive to alteration and understanding. In this study, in order to obtain CAMs of our cerebral infarction MRI images more accurately and efficiently, we test and compare several traditional standard classification networks, including AlexNet, GoogLeNet, VGG, ResNet, and DenseNet. Among them, AlexNet, designed in 2012, has a more complex architecture [59]. ReLu is used instead of sigmoid as the activation function to solve the gradient dispersion problem of sigmoid in deep networks. Dropout is used in training to ignore some neurons to avoid overfitting randomly. GoogLeNet [60] adopts different convolution kernels to obtain receptive fields, approximating the optimal sparse structure by building a dense block structure.

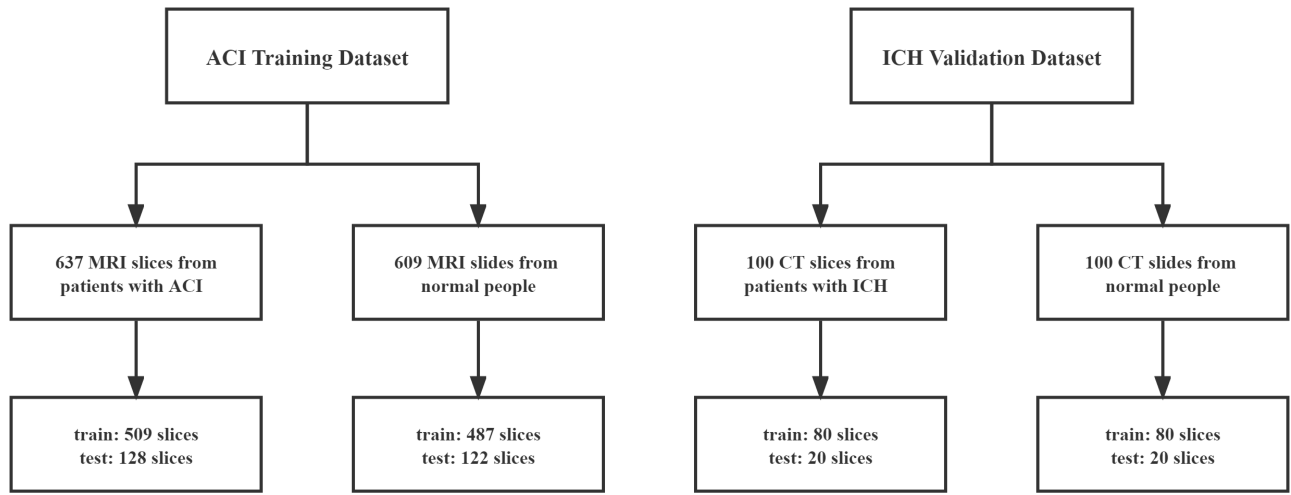

Fig. 1. Composition of ACI training dataset and ICH validation dataset.

VGG [61] conducts a comprehensive evaluation of the depth dependence of the network using an architecture with small convolutional filters, which shows that significant improvements can be achieved with deeper convolutional layers. The deep Residual Network (ResNet) [62] conducts residual learning to deeper networks without additional parameters. This trick allows the network to learn identity mappings from previous layers, so that deeper models perform at least no worse than shallower models. DenseNet [52] introduces connections from one layer to all its subsequent layers in a feed-forward fashion, where the layers are narrow and alleviate the problem of vanishing gradients.

### C. Multi-Scale Reasoning of CAMs

We can get the most discriminative target area for a well-trained classification network through CAMs [33]. The classification network's full connection and softmax layers are replaced by global average pooling (GAP). The average value of all pixels in the feature map returns the whole feature map. To be more specific, let $A_{ij}^k$ represents the active unit $k$ at the spatial position $(i,j)$ of the feature map before the GAP layer and each feature map has a corresponding weight $w_k^c$. Z represents the size of the input image, where a and b represent length and width respectively. Then, we can obtain an activation map $Cam^c$ and class score $Y^c$ of the corresponding class $c$ as follows:

$$Cam^c = \sum_k w_k^c \times A^k \quad (1)$$

$$Y^c = \sum_k w_k^c \frac{1}{Z} \sum_i \sum_j A_{ij}^k, Z = a \times b \quad (2)$$

People use the original CAMs to conduct the follow-up operation. However, the problem that can not be ignored is that the original CAMs can only identify the image regions most relevant to the particular category, which is highly dissimilar to ground truth. Applying different rescaling transformations on the input images will not be able to obtain the same transformations on the generated CAMs, which is mainly due to the supervision gap between fully and weakly supervised semantic segmentation. Inspired by this inequivalence, we propose a method of multi-scale reasoning to refine CAMs. In detail, for an input image $i$, we sample it $m$ times by setting different sampling rates. Then, $S_{i,c}^j$ represents the CAM of category $c$ corresponding to the scale $j$ of the input image $i$. Finally, we can obtain the fused CAM $S_c'$ as following:

$$S_c' = \sum_j^m \frac{S_{i,c}^j}{m} \quad (3)$$

### D. Medical Image Segmentation with Mixed-UNet

Most of the current semantic segmentation methods adopt an encoder-decoder architecture based on deep learning, such as FCN [1], UNet [2], Attention UNet [3], SegNet [4], DeepLab [7], and PSPNet [6]. The deep semantic information is extracted in the encoding phase and then is mapped to the given category region in the decoding phase. However, these methods use one branch in the decoding stage, affecting the performance of weakly supervised semantic segmentation because of inaccurate labels obtained from CAMs. We, therefore, put forward a new segmentation model for weakly supervised semantic segmentation as illustrated in Figure 2, namely Mixed-UNet.

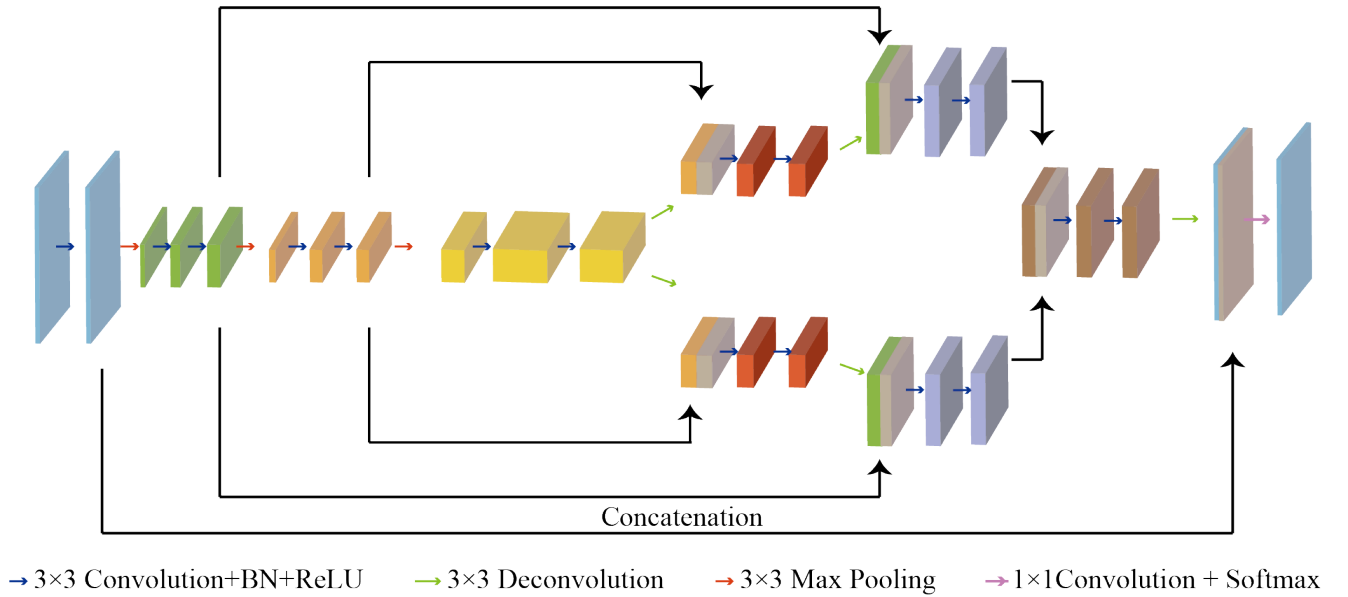

Fig. 2. An illustration of the Mixed-UNet architecture. The coding phase consists of the repeated application of two 3x3 convolutional layers and one 3x3 max pooling operation. Besides, each 3x3 convolutional layer is followed by a BN and a ReLU. Correspondingly, in the decoding phase, each step upsamples the feature map followed by a 3x3 deconvolution to halve the number of feature channels. The feature maps obtained by deconvolution in upsampling are concatenated with the corresponding cropped feature maps in downsampling. The fused feature maps are subjected to the same convolution operation. The final layer that consists of a 1x1 convolutional layer and a softmax layer is used to map each 64 component feature vector to the desired number of classes.

There are three downsampling in the coding stage for the segmentation because the lesion area is always tiny and scattered, which is one layer less than U-Net. It can save more spatial details and reduce computational complexity, contributing to high efficiency, which is essential in medical images processing. Unlike double U-Net [63], cascaded V-Net [64], where two networks are connected in a cascade way, our method combines two U-Nets in a mixed way. The coding phase shares a feature extractor and is divided into two parallel branches in the decoding phase, which have the same structure. In detail, the downsampling follows the typical architecture of a convolutional network. It consists of the repeated application of two 3x3 convolutional layers and one 3x3 max pooling operation. Besides, each 3x3 convolutional layer is followed by a Batch Normalization layer (BN) and a rectified linear unit (ReLU). In particular, each downsampling step in Figure 2 is distinguished by blocks of different colors, and we double the number of feature channels in each step. Correspondingly, in the decoding phase, each step upsamples the feature map followed by a 3x3 deconvolution to halve the number of feature channels. Then, the feature maps obtained by deconvolution in upsampling are concatenated with the corresponding cropped feature maps in downsampling. The fused feature maps are subjected to the same convolution operation. The cropping is necessary due to the loss of border pixels in every convolution. And after the fuse of the features extracted from the two branches, the final layer that consists of a 1x1 convolutional layer and a softmax layer is used to map each 64 component feature vector to the desired number of classes, obtaining the final segmentation result. Non-linear activation as the softmax layer is added to improve the expression ability of the network.

For an input image $x_i$, $ef$ represents the semantic feature information extracted in encoding stage and $df_i$ represents the feature extracted by the $ith$ branch in the decoding phase. $y_i$ represents the final predicted segmentation result.

We can use the following formulas to describe the overall operations:

$$ef = encoding(x_i) \qquad (4)$$

$$df_1 = decoding(ef) \qquad (5)$$

$$df_2 = decoding(ef) \qquad (6)$$

$$y_i = softmax(conv(concat(df_1, df_2))) \qquad (7)$$

To train the proposed Mixed-UNet model, we combine seeding loss [59] using Refined_mask and cross-entropy loss function using CRF_mask.

Supposing $X$ is the input image, $T$ is the category of classification, $Y$ is the output of the model, and $u$ represents any position in the space, $S_c$ is a set of locations that are labeled with class $c$ by the weak localization procedure. We define the following loss function:

$$L_{seed}(Y,T,S_c) = -\frac{1}{\sum_{c\in T}\|S_c\|}\sum_{c\in T}\sum_{u\in S_c}\log Y_{u,c} \qquad (8)$$

$$L_{CE}(Y,T) = -\sum_{i=1}^{T} T_i \log Y_i \qquad (9)$$

$$L_{loss} = L_{seed}(f(X),T,S_c) + L_{CE}(Y,T) \qquad (10)$$

## IV. EXPERIMENT AND RESULT

### A. Classification of medical image

We use the CNN models to generate CAMs and test the classification performance of ACI with popular CNN models of AlexNet, GoogLeNet, VGG, ResNet, and DenseNet, respectively.

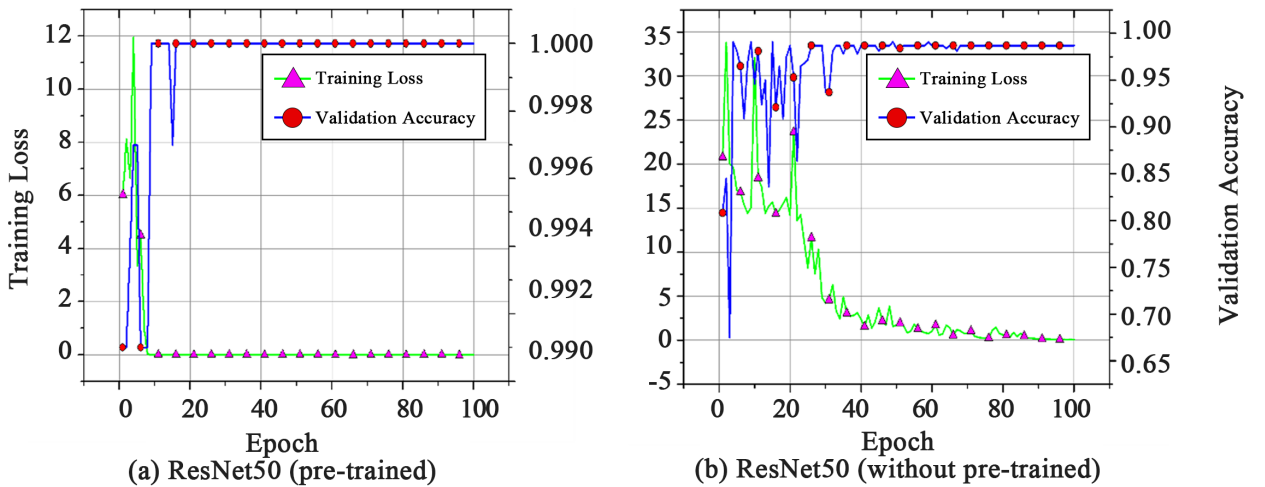

Fig. 3. Training loss and validation accuracy of ResNet50 (pre-trained) and ResNet50 (without pre-trained).

Table I reports the diagnosis classification of the diagnosis of acute cerebral infarction from brain MRI. While all the methods achieve promising performance, ResNet50 has fewer parameters and faster convergence speed. The pre-trained ResNet50 has higher accuracy and faster convergence speed. Therefore, we choose pre-trained ResNet50 as our architecture.

TABLE I
CLASSIFICATION RESULTS FOR ACI AND NORMAL MRI SLICES

| Model | Accuracy |
|---|---|
| AlexNet | 0.99 |
| GoogLeNet | 0.99 |
| VGG | 1.00 |
| ResNet50(without pre-trained) | 0.987 |
| ResNet50(pre-trained) | 1.00 |
| DenseNet | 1.00 |

We compare the performance of ResNet50 pre-trained with that without pre-trained, as illustrated in Figure 3. The pre-trained ResNet50 has higher accuracy and faster convergence speed, which is selected to be used in our architecture for the following CAMs generation.

### B. Results of Multi-scale Inference

After training, we can get the CAMs through the classification model to locate the most significant area of the target. We sample the original image four times with different sizes, and the sampling rates are 0.5 (scale_2), 1.0 (origin_CAM), 1.5 (scale_3) and 2.0 (scale_4), respectively. Figure 4 shows the result of multi-scale inference, and refined CAMs refer to CAMs after fusing four scales of CAMs. The original CAMs can only locate the general area of the lesion, but it differs significantly from the ground truth in size and boundary. In addition, for multiple scattered lesions, origin CAMs can only identify the most apparent lesions while ignoring other difficultly identified lesions. The shape and size of the refined CAMs are more consistent with the lesion area, which can find multiple lesion areas. Then, we normalize the refined CAMs. Let $x_c$ be any position of CAM of class $c$. By using the following normalization formula, $x_c$ can be normalized into a number between [0,1]:

$$x'_c = \frac{x_c - x_{c,min}}{x_{c,max} - x_{c,min}} \qquad (11)$$

$x_{c,max}$ and $x_{c,min}$ refer to the maximum and minimum value of CAM of different class $c$. We take the following threshold segmentation method to obtain the pseudo masks used for semantic segmentation:

$$f'(x,y) = \begin{cases} 1, & f(x,y) \geq T \\ 0, & f(x,y) < T \end{cases} \qquad (12)$$

Where T is the threshold depending on the lesions, we set T to be 0.35. Here, the threshold needs a special explanation. When the lesion range is large, such as the lesion of ACI, we need to adjust the threshold value to a lower level to obtain a better segmentation effect. A higher threshold is required for small focal areas, such as ICH's small and scattered bleeding sites. $f(x,y)$ refers to the pixels whose spatial position is (x, y) in refined CAMs. In order to better estimate the boundary of lesions, we use the dense CRF model [7] to optimize the pseudo masks, which will assign the same label to the pixels with similar features in the image when inputting the pseudo masks and the original image as illustrated in Figure 5. The original mask is very rough, and if there are multiple lesions, only one lesion can be detected. Refined_mask after multi-scale inference are consistent with the original lesion area in size and shape but can detect additional lesion areas. Based on Refined_mask, CRF_mask optimized by dense CRF can produce competitive results closer to the ground truth. Table II shows the numerical comparison of different masks. CRF_mask has the highest predicted segmentation (PA) and mean intersection over union (MIoU).

TABLE II
NUMERICAL COMPARISON RESULTS OF DIFFERENT MASKS

| Mask | PA | MIoU |
|---|---|---|
| Origin_mask | 0.989 | 0.621 |
| Refined_mask | 0.991 | 0.652 |
| CRF_mask | 0.993 | 0.726 |

### C. Mixed-UNet Construction

For semantic segmentation, 60% of data is used for training, 20% of data is used for model selection and hyper-parameters adjustment, and the remaining 20% is used for testing. We compare our designed Mixed-UNet algorithm with state-of-the-art methods: FCN [1], U-Net [2], and DeepLab V3 [7]. The model is trained on 4 NVIDIA TITAN X GPUs with batch size 8 for 100 epochs. The Mixed-UNet is first trained using the Adam optimizer [65] with an initial learning rate ($lr_{init}$) of 1e$^{-3}$. The practical learning rate follows a polynomial decay, to be zero until the max iteration.

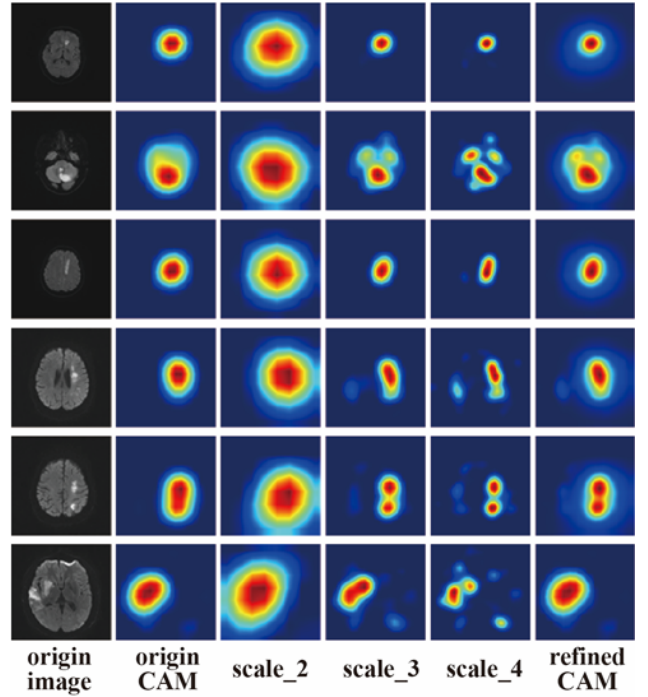

Fig. 4. Visualization of multi-scale CAMs inference.

888

The poly policy follows (13) with $\gamma = 0.9$ for decay:

$$lr_{itr} = lr_{init}\left(1 - \frac{itr}{max - itr}\right)^{\gamma} \quad (13)$$

Due to the limited training images, we employ random flipping (up-down or left-right), random rotation (one of ± 25, 90, 180, or 270 degrees), and random Gaussian noise addition ($\sigma$ from 0.3 to 0.7) to augment data. We use $\mathcal{L}_2$ regularization on the network parameters with weight $\lambda = 10^{-4}$. The original MRI slices have a different number of pixels spatially, and we resize them to 256 * 256. The qualitative segmentation results of Mixed-UNet on samples including acute cerebral infarction are shown in Figure 6 Mixed-UNet obtains reasonable predictions due to the design of parallel branch structure, which can extract more information than a single branch.

TABLE III
THE QUANTITATIVE COMPARISON OF DIFFERENT MODELS

| Model | PA | MIoU | Dice |
|---|---|---|---|
| FCN | 0.991 | 0.602 | 0.409 |
| UNet | 0.991 | 0.615 | 0.487 |
| DeepLab V3 | 0.991 | 0.656 | 0.529 |
| Mixed-UNet | 0.992 | 0.645 | 0.560 |
| Mixed-UNet (+CRF) | 0.992 | 0.674 | 0.543 |

The quantitative results in Table III show that all the deep architectures have high PA and low Dice. One reason for the overall poor performance is that the focus areas occupy a small part of the image while the background occupies the majority space of the image.

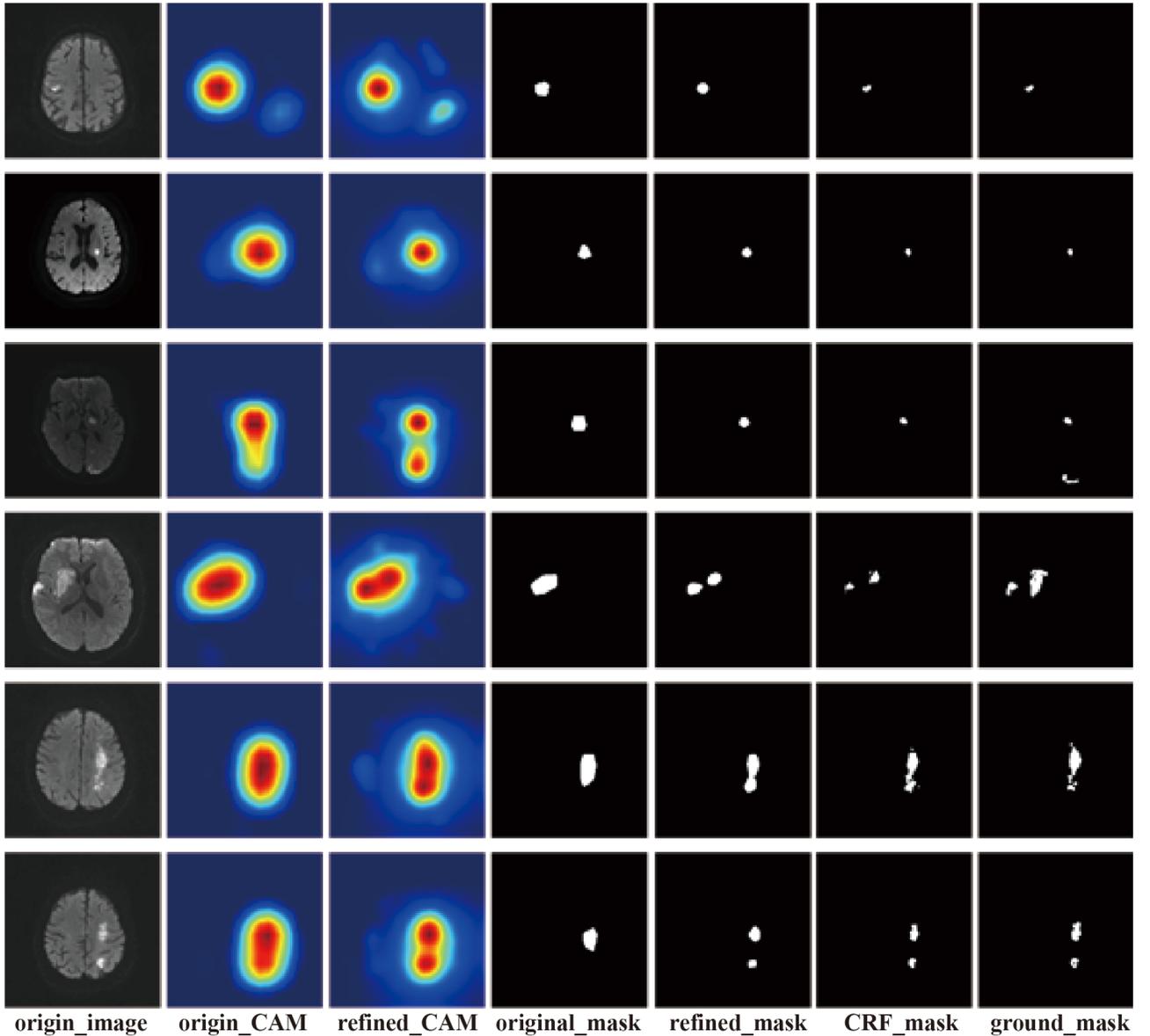

**origin_image  origin_CAM  refined_CAM  original_mask  refined_mask  CRF_mask  ground_mask**

Fig. 5. Representative CAMs depicting the lesion boundary with different optimization models.

A small error in the background will significantly influence the segmentation results. Our model achieves better optimal results, except MIoU is slightly lower than DeepLab V3. When we join CRF_mask, using Refined_mask and CRF_mask to train the network simultaneously, we find that MIoU increases by 3.1%, but Dice decreases by 1.7%. More optimal settings should exist theoretically, but the grid search is required, which is computational expensive given the considerable inference time for dense-CRFs. Table IV compares several different deep learning models regarding parameter quantity, memory consumption, and computation time spent on segmenting an image. Compared with other methods, Mixed-UNet has the least number of parameters and minor memory consumption, which is a quarter of Deeplab V3. However, in the case of fewer parameters, Mixed-UNet takes the longest time to infer an image, mainly due to the structure of two parallel branches.

The parallel structure integrates the feature extracted from two branches and pinpoints more tiny details that a single branch can not find, whose cost is the computation time needed to mine the subtle details.

TABLE IV
A COMPARISON OF COMPUTATIONAL TIME AND HARDWARE RESOURCES REQUIRED FOR VARIOUS DEEP ARCHITECTURES

| Model | # of Trainable Parameters | Memory | Time(s) |
|---|---|---|---|
| FCN (VGG16) | 18,643,713 | 73M | 0.421 |
| U-Net | 31,043,521 | 121M | 0.861 |
| DeepLab V3 | 39,633,729 | 155M | 0.936 |
| Mixed-UNet | 10,303,105 | 50M | 1.153 |

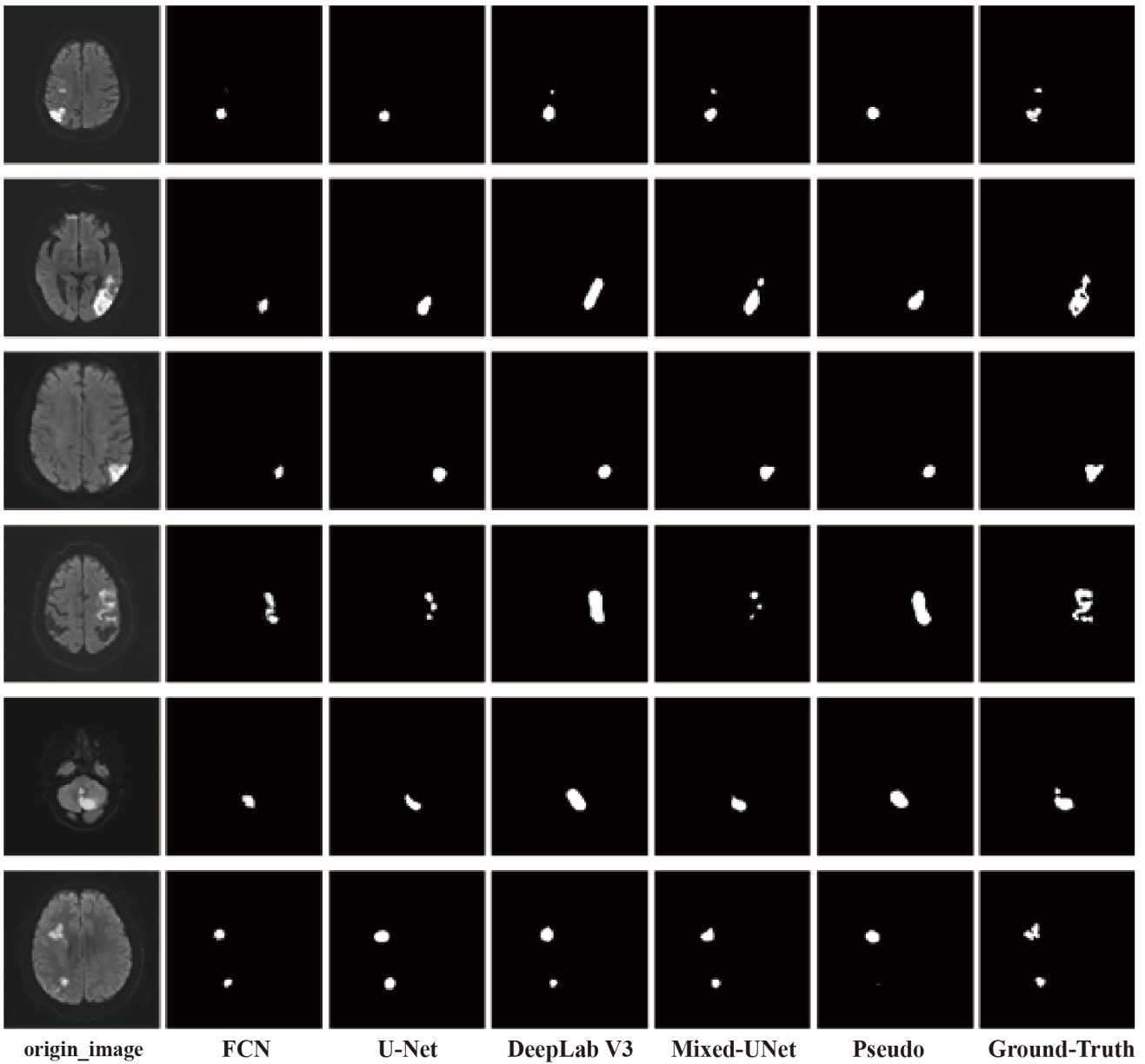

Fig. 6. Representative segmentation results with Mixed-UNet and other methods.

## D. Validation of the Mixed-UNet

We further evaluate our model on the ICH dataset and set the segmentation threshold to 0.7 due to the tiny lesions of the ICH dataset. The results show that our method can significantly improve the weakly supervised classification performance by providing refined pixel-level labels. A dense CRF model with refined CAMs shows a remarkable enhancement of depicting the boundary of lesions as illustrated in Figure 7. The original mask is limited to detecting partial lesions and cannot cover the lesions in massive bleeding images. Refined_mask can capture the rough shape of lesion areas. On the premise of accurate location of Refined_mask and utilization of dense CRF, CRF_mask demonstrates an ability to segment details of lesions. The public dataset does not include the expert ground label, which is labeled by the experienced expert for metric evaluation calculation. CRF_mask has the highest PA (0.9831), MioU (0.7706), and Dice (0.6907) among the output masks.

## V. DISCUSSION AND CONCLUSION

In this study, we present a weekly supervised semantic segmentation framework that integrates the image classification and segmentation of cerebral infarction. We propose multi-scale inference to optimize origin CAMs. Meanwhile, we develop a new semantic segmentation network named Mixed-UNet, adopting two paralleled branches in the upsampling phase and testing the performance using Refined_Mask and CRF_Mask as the supervision. The experimental results show that our models perform better due to the multi-scale inference and two paralleled branches. The limitation of our method is the supervised way of training, and it does not dig out the information except the label. In this case, when the label is inaccurate, label dependence is worth further investigation. The model is currently used to pinpoint the brain imaging lesion, which can definitely be extended to other imaging modalities and various lesion detection.

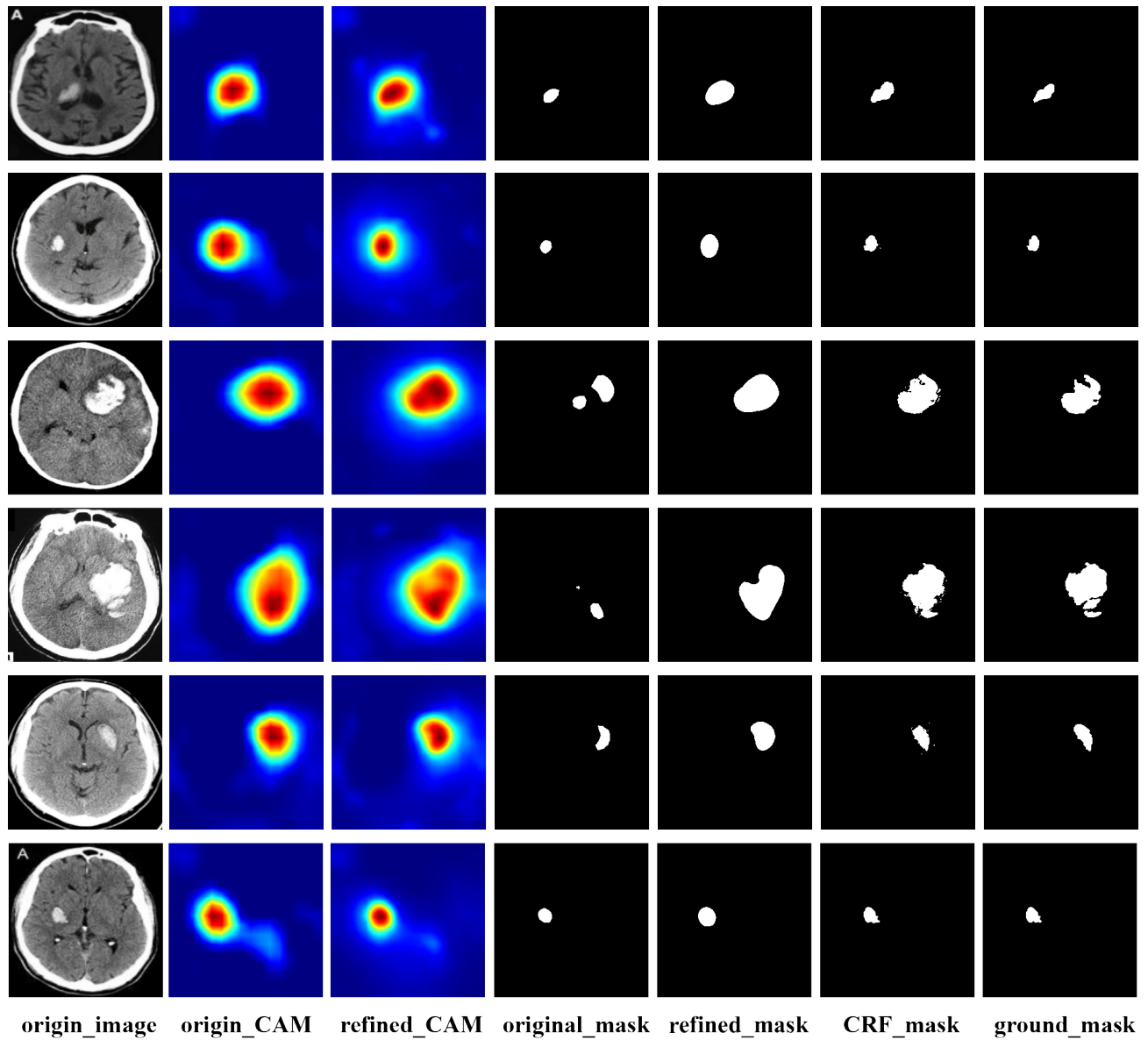

Fig. 7. The validation of the Mixed-UNet and other methods.

## REFERENCES

[1] J. Long, E. Shelhamer, and T. Darrell, "Fully convolutional networks for semantic segmentation," in *Proc. IEEE Conf. Comput. Vis. Pattern Recognit. (CVPR)*, 2015, pp. 3431-3440.

[2] O. Ronneberger, P. Fischer, and T. Brox, "U-net: Convolutional networks for biomedical image segmentation," in *Med. Image Comput. Comput. Assist. Interv. (MICCAI)*, 2015, pp. 234-241.

[3] O. Oktay, J. Schlemper, L. L. Folgoc *et al.*, "Attention U-net: Learning where to look for the pancreas," 2018, *arXiv:1804.03999*.

[4] V. Badrinarayanan, A. Kendall, and R. Cipolla, "Segnet: A deep convolutional encoder-decoder architecture for image segmentation," *IEEE Trans. Pattern Anal. Mach. Intell.*, vol. 39, no. 12, pp. 2481-2495, 2017.

[5] F. Yu, and V. Koltun, "Multi-scale context aggregation by dilated convolutions," 2015, *arXiv:1511.07122*.

[6] H. Zhao, J. Shi, X. Qi *et al.*, "Pyramid scene parsing network," in *Proc. IEEE Conf. Comput. Vis. Pattern Recognit. (CVPR)*, 2017, pp. 2881-2890.

[7] L.-C. Chen, G. Papandreou, I. Kokkinos *et al.*, "Deeplab: Semantic image segmentation with deep convolutional nets, atrous convolution, and fully connected crfs," *IEEE Trans. Pattern Anal. Mach. Intell.*, vol. 40, no. 4, pp. 834-848, 2017.

[8] A. S. Lundervold, and A. Lundervold, "An overview of deep learning in medical imaging focusing on MRI," *Z. Med. Phys.*, vol. 29, no. 2, pp. 102-127, 2019.

[9] S. A. Taghanaki, K. Abhishek, J. P. Cohen *et al.*, "Deep semantic segmentation of natural and medical images: a review," *Artif. Intell. Rev.*, vol. 54, no. 1, pp. 137-178, 2021.

[10] R. Yang, and Y. Yu, "Artificial Convolutional Neural Network in Object Detection and Semantic Segmentation for Medical Imaging Analysis," *Front. Oncol.*, vol. 11, pp. 638182, 2021.

[11] J. S. Kim, and L. R. Caplan, "Clinical Stroke Syndromes," *Front Neurol Neurosci*, vol. 40, pp. 72-92, 2016.

[12] S. Warach, J. Gaa, B. Siewert *et al.*, "Acute human stroke studied by whole brain echo planar diffusion-weighted magnetic resonance imaging," *Ann. Neurol.*, vol. 37, no. 2, pp. 231-241, 1995.

[13] N. Pourian, S. Karthikeyan, and B. S. Manjunath, "Weakly supervised graph based semantic segmentation by learning communities of image-parts," in *Proc. IEEE Int. Conf. Comput. Vis. (ICCV)*, 2015, pp. 1359-1367.

[14] J. Xu, A. G. Schwing, and R. Urtasun, "Learning to segment under various forms of weak supervision," in *Proc. IEEE Conf. Comput. Vis. Pattern Recognit. (CVPR)*, 2015, pp. 3781-3790.

[15] A. Bearman, O. Russakovsky, V. Ferrari *et al.*, "What's the point: Semantic segmentation with point supervision," 2016, *arxiv:1506.02106*.

[16] G. Papandreou, L.-C. Chen, K. P. Murphy *et al.*, "Weakly-and semi-supervised learning of a deep convolutional network for semantic image segmentation," in *Proc. IEEE Int. Conf. Comput. Vis. (ICCV)*, 2015, pp. 1742-1750.

[17] X. Qi, J. Shi, S. Liu *et al.*, "Semantic segmentation with object clique potential," in *Proc. IEEE Int. Conf. Comput. Vis. (ICCV)*, 2015, pp. 2587-2595.

[18] D. Pathak, E. Shelhamer, J. Long *et al.*, "Fully convolutional multi-class multiple instance learning," 2014, *arXiv:1412.7144*.

[19] P. O. Pinheiro, and R. Collobert, "From image-level to pixel-level labeling with convolutional networks," in *Proc. IEEE Conf. Comput. Vis. Pattern Recognit. (CVPR)*, 2015, pp. 1713-1721.

[20] D. Pathak, P. Krahenbuhl, and T. Darrell, "Constrained convolutional neural networks for weakly supervised segmentation," in *Proc. IEEE Int. Conf. Comput. Vis. (ICCV)*, 2015, pp. 1796-1804.

[21] J. Dai, K. He, and J. Sun, "Boxsup: Exploiting bounding boxes to supervise convolutional networks for semantic segmentation," in *Proc. IEEE Int. Conf. Comput. Vis. (ICCV)*, 2015, pp. 1635-1643.

[22] C. Song, Y. Huang, W. Ouyang *et al.*, "Box-driven class-wise region masking and filling rate guided loss for weakly supervised semantic segmentation," in *Proc. IEEE Conf. Comput. Vis. Pattern Recognit. (CVPR)*, 2019, pp. 3136-3145.

[23] D. Lin, J. Dai, J. Jia *et al.*, "Scribblesup: Scribble-supervised convolutional networks for semantic segmentation," in *Proc. IEEE Conf. Comput. Vis. Pattern Recognit. (CVPR)*, 2016, pp. 3159-3167.

[24] R. Briq, M. Moeller, and J. Gall, "Convolutional simplex projection network (CSPN) for weakly supervised semantic segmentation," 2018, *arXiv:1807.09169*.

[25] X. Wang, S. You, X. Li *et al.*, "Weakly-supervised semantic segmentation by iteratively mining common object features," in *Proc. IEEE Conf. Comput. Vis. Pattern Recognit. (CVPR)*, 2018, pp. 1354-1362.

[26] Y. Wei, H. Xiao, H. Shi *et al.*, "Revisiting dilated convolution: A simple approach for weakly-and semi-supervised semantic segmentation," in *Proc. IEEE Conf. Comput. Vis. Pattern Recognit. (CVPR)*, 2018, pp. 7268-7277.

[27] Z. Huang, X. Wang, J. Wang *et al.*, "Weakly-supervised semantic segmentation network with deep seeded region growing," in *Proc. IEEE Conf. Comput. Vis. Pattern Recognit. (CVPR)*, 2018, pp. 7014-7023.

[28] X. Zhang, Y. Wei, J. Feng *et al.*, "Adversarial complementary learning for weakly supervised object localization," in *Proc. IEEE Conf. Comput. Vis. Pattern Recognit. (CVPR)*, 2018, pp. 1325-1334.

[29] P. Vernaza, and M. Chandraker, "Learning random-walk label propagation for weakly-supervised semantic segmentation," in *Proc. IEEE Conf. Comput. Vis. Pattern Recognit. (CVPR)*, 2017, pp. 7158-7166.

[30] A. Roy, and S. Todorovic, "Combining bottom-up, top-down, and smoothness cues for weakly supervised image segmentation," in *Proc. IEEE Conf. Comput. Vis. Pattern Recognit. (CVPR)*, 2017, pp. 3529-3538.

[31] S. Kwak, S. Hong, and B. Han, "Weakly supervised semantic segmentation using superpixel pooling network," in *Proc. Conf. AAAI. Artif. Intell.*, 2017.

[32] J. Ahn, and S. Kwak, "Learning pixel-level semantic affinity with image-level supervision for weakly supervised semantic segmentation," in *Proc. IEEE Conf. Comput. Vis. Pattern Recognit. (CVPR)*, 2018, pp. 4981-4990.

[33] B. Zhou, A. Khosla, A. Lapedriza *et al.*, "Learning deep features for discriminative localization," in *Proc. IEEE Conf. Comput. Vis. Pattern Recognit. (CVPR)*, 2016, pp. 2921-2929.

[34] P. T. Jiang, C. B. Zhang, Q. Hou *et al.*, "LayerCAM: Exploring Hierarchical Class Activation Maps for Localization," *IEEE Trans. Image Process.*, vol. 30, pp. 5875-5888, 2021.

[35] K. Li, Z. Wu, K.-C. Peng *et al.*, "Tell me where to look: Guided attention inference network," in *Proc. IEEE Conf. Comput. Vis. Pattern Recognit. (CVPR)*, 2018, pp. 9215-9223.

[36] B. Wang, C. Yuan, B. Li *et al.*, "Multi-scale low-discriminative feature reactivation for weakly supervised object localization," *IEEE Trans. Image Process.*, vol. 30, pp. 6050-6065, 2021.

[37] P. Krähenbühl, and V. Koltun, "Efficient inference in fully connected crfs with gaussian edge potentials," *Adv. Neural Inf. Process. Syst.*, vol. 24, 2011.

[38] B. Liang, Y. Liu, L. He *et al.*, "Weakly supervised semantic segmentation based on deep learning," in *Proc. IASTED Int. Conf. Model. Identif. Control (ICMIC)*, Singapore, 2020, pp. 455-464.

[39] A. Vezhnevets, V. Ferrari, and J. M. Buhmann, "Weakly supervised semantic segmentation with a multi-image model," in *Proc. IEEE Int. Conf. Comput. Vis. (ICCV)*, 2011, pp. 643-650.

[40] Y. Wei, W. Xia, M. Lin *et al.*, "HCP: A Flexible CNN Framework for Multi-Label Image Classification," *IEEE Trans. Pattern Anal. Mach. Intell.*, vol. 38, no. 9, pp. 1901-1907, 2016.

[41] Y. Wei, X. Liang, Y. Chen *et al.*, "Learning to segment with image-level annotations," *Pattern Recognit.*, vol. 59, pp. 234-244, 2016.

[42] A. Kolesnikov, and C. H. Lampert, "Seed, expand and constrain: Three principles for weakly-supervised image segmentation," in *Proc. ECCV*, Cham, 2016, pp. 695-711.

[43] R. R. Selvaraju, M. Cogswell, A. Das *et al.*, "Grad-cam: Visual explanations from deep networks via gradient-based localization," in *Proc. IEEE Int. Conf. Comput. Vis. (ICCV)*, 2017, pp. 618-626.

[44] A. Chattopadhay, A. Sarkar, P. Howlader *et al.*, "Grad-cam++: Generalized gradient-based visual explanations for deep convolutional networks," in *Proc. IEEE Winter Conf. Appl. Comput. Vis. (WACV)*, 2018, pp. 839-847.

[45] H. Wang, Z. Wang, M. Du *et al.*, "Score-CAM: Score-weighted visual explanations for convolutional neural networks," in *Proc. IEEE Conf. Comput. Vis. Pattern Recognit. (CVPR)*, 2020, pp. 24-25.

[46] M. Oquab, L. Bottou, I. Laptev *et al.*, "Is object localization for free?-weakly-supervised learning with convolutional neural networks," in *Proc. IEEE Conf. Comput. Vis. Pattern Recognit. (CVPR)*, 2015, pp. 685-694.

[47] J. Choe, and H. Shim, "Attention-based dropout layer for weakly supervised object localization," in *Proc. IEEE Conf. Comput. Vis. Pattern Recognit. (CVPR)*, 2019, pp. 2219-2228.

[48] S. Lee, J. Lee, J. Lee *et al.*, "Robust tumor localization with pyramid grad-cam," 2018, *arXiv:1805.11393*.

[49] G. Lin, A. Milan, C. Shen *et al*., "Refinenet: Multi-path refinement networks for high-resolution semantic segmentation," in *Proc. IEEE Conf. Comput. Vis. Pattern Recognit. (CVPR)*, 2017, pp. 1925-1934.
[50] J. Shen, X. Tang, X. Dong *et al*., "Visual object tracking by hierarchical attention siamese network," *IEEE Trans. Cybern.,* vol. 50, no. 7, pp. 3068-3080, 2019.
[51] S. Ioffe, and C. Szegedy, "Batch normalization: Accelerating deep network training by reducing internal covariate shift," in *Proc. ICML*, 2015, pp. 448-456.
[52] G. Huang, Z. Liu, L. Van Der Maaten *et al*., "Densely connected convolutional networks," in *Proc. IEEE Conf. Comput. Vis. Pattern Recognit. (CVPR)*, 2017, pp. 4700-4708.
[53] L.-C. Chen, Y. Zhu, G. Papandreou *et al*., "Encoder-decoder with atrous separable convolution for semantic image segmentation," in *Proc. ECCV*, 2018, pp. 801-818.
[54] L.-C. Chen, G. Papandreou, F. Schroff *et al*., "Rethinking atrous convolution for semantic image segmentation," 2017, *arXiv:1706.05587*.
[55] Y. Weng, T. Zhou, Y. Li *et al*., "Nas-unet: Neural architecture search for medical image segmentation," *IEEE Access,* vol. 7, pp. 44247-44257, 2019.
[56] R. Brügger, C. F. Baumgartner, and E. Konukoglu, "A partially reversible u-net for memory-efficient volumetric image segmentation," in *Med. Image Comput. Comput. Assist. Interv. (MICCAI)*, 2019, pp. 429-437.
[57] Y. Wu, and K. He, "Group normalization," in *Proc. ECCV*, 2018, pp. 3-19.
[58] L. R. Bonta, and N. U. Kiran, "Efficient segmentation of medical images using dilated residual networks," in *Computer Aided Intervention and Diagnostics in Clinical and Medical Images*: Springer, 2019, pp. 39-47.
[59] A. Krizhevsky, I. Sutskever, and G. E. Hinton, "Imagenet classification with deep convolutional neural networks," *Adv. Neural Inf. Process. Syst.,* vol. 25, pp. 1097-1105, 2012.
[60] C. Szegedy, W. Liu, Y. Jia *et al*., "Going deeper with convolutions," in *Proc. IEEE Conf. Comput. Vis. Pattern Recognit. (CVPR)*, 2015, pp. 1-9.
[61] K. Simonyan, and A. Zisserman, "Very deep convolutional networks for large-scale image recognition," 2014, *arXiv:1409.1556*.
[62] K. He, X. Zhang, S. Ren *et al*., "Deep residual learning for image recognition," in *Proc. IEEE Conf. Comput. Vis. Pattern Recognit. (CVPR)*, 2016, pp. 770-778.
[63] D. D. Pham, G. Dovletov, S. Warwas *et al*., "Deep segmentation refinement with result-dependent learning," in *Bildverarbeitung für die Medizin*: Springer, 2019, pp. 49-54.
[64] A. Casamitjana, M. Catà, I. Sánchez *et al*., "Cascaded V-Net using ROI masks for brain tumor segmentation," in *Proc. MICCAI BrainLes. Workshops*, 2017, pp. 381-391.
[65] D. P. Kingma, and J. Ba, "Adam: A method for stochastic optimization," 2014, *arXiv:1412.6980*.